\documentclass[pra,twocolumn,superscriptaddress,amsmath,amssymb]{revtex4-1}
\usepackage{graphicx} 
\usepackage{dcolumn}  
\usepackage{bm}       
\usepackage{amsfonts}
\usepackage{color}

\usepackage{soul,xcolor}
\setstcolor{red}
\begin{document}
\title{Imaging charge migration in chiral molecules using time-resolved x-ray diffraction}

\author{Sucharita Giri}
\email[]{sucharitagiri@iitb.ac.in}
\affiliation{%
Department of Physics, Indian Institute of Technology Bombay,
            Powai, Mumbai 400076  India}
            
\author{Jean Christophe Tremblay}
\email[]{jean-christophe.tremblay@univ-lorraine.fr}
\affiliation{%
Laboratoire de Physique et Chimie Th\'eoriques, 
CNRS-Universit\'e de Lorraine, UMR 7019, ICPM, 1 Bd Arago, 57070 Metz, France}

\author{Gopal Dixit}
\email[]{gdixit@phy.iitb.ac.in}
\affiliation{%
Department of Physics, Indian Institute of Technology Bombay,
            Powai, Mumbai 400076  India}

\date{\today}



\begin{abstract}
Four-dimensional imaging of charge migration is  crucial to the understanding of several ubiquitous processes in nature. The present work focuses on imaging of charge migration in an oriented epoxypropane: a chiral molecule. 
A linearly  polarized  pulse is used to induce the charge migration, which is imaged by time-resolved x-ray diffraction.   
It is found that the total time-resolved diffraction signals  
are significantly different for both enantiomers.    
Furthermore, a connection  between time-resolved x-ray diffraction and the electronic continuity equation is discussed by analyzing the time-dependent  diffraction signal and  the time derivative of the  total  electron density in the momentum space. 
\end{abstract}

\maketitle
\section{Introduction}

X-ray diffraction is the method of choice when it comes to  
extracting the real-space structure of solids and molecules with atomic-scale spatial resolution~\cite{Nielsen}.  
Thanks to technological  advances that have made it possible to 
generate intense ultrashort pulses from novel x-ray source: x-ray free-electron lasers 
(XFELs)~\cite{ishikawa2012, pellegrini2016physics, emma2}. 
X-ray pulses with pulse duration of a few femtoseconds are routinely employed to perform various kinds of experiments at different 
XFEL sources. 
Moreover, generation of attosecond x-ray pulses was demonstrated experimentally  in recent years~\cite{hartmann2018attosecond, bucksbaum2019attoXR}.
These ultrashort x-ray pulses have enabled extension of  
static x-ray diffraction into the time domain with 
exceptional temporal resolution~\cite{lindroth2019challenges}.  
Time-resolved x-ray diffraction  (TRXD) within a pump-probe configuration  is an emerging 
method to obtain snapshots of a temporarily 
evolving electronic charge distribution in matter with spatiotemporal resolution on atomic and electronic scales~\cite{peplow2017next}.  
An electronic movie of  ultrafast charge migration can be made by  
putting the recorded snapshots together as a function of pump-probe delay time~\cite{dixit2012, vrakking2012}.  

In this work, we investigate imaging of  pump-induced charge migration in a chiral molecule using TRXD. 
Oriented epoxypropane (1,2-propylene oxide) is used as a prototypical system.
The notion of chirality is omnipresent in nature~\cite{barrett2014spontaneous, gal2012discovery}. 
Understanding the chirality is essential for a broad range of sciences including origin of homochirality on earth.
A molecule without inversion symmetry or a symmetry plane which is not superimposable with its own mirror image is known as a chiral molecule~\cite{cahn1966specification}. 
A pair of such chiral molecules is known as an enantiomers. 
Enantiomers exhibit identical physical properties, but show a strong enantiomeric preference during chemical and biological reactions. 
Discerning the enantiomeric excess and handedness of chiral molecules is crucial  in chemistry, biology and pharmaceuticals. 

Confirming the handedness of chiral molecules by determining their spatial arrangement is 
known as absolute configuration determination~\cite{Nielsen, bijvoet1951determination, santoro2020absolute}. 
X-ray crystallography is one of the most reliable methods to determine  
the absolute configuration~\cite{flack1999absolute}. 
In recent years, other methods such as microwave spectroscopy~\cite{patterson2013enantiomer, eibenberger2017enantiomer},  Coulomb-explosion imaging~\cite{pitzer2013direct, herwig2013imaging},  high-harmonic generation~\cite{cireasa2015probing, wang2017high, neufeld2019ultrasensitive, baykusheva2019real}, vibrational circular dichroism~\cite{fuse2019vibrational, magyarfalvi2011vibrational}, and photoelectron circular dichroism~\cite{bowering2001asymmetry, lux2012circular, dreissigacker2014photoelectron, beaulieu2017attosecond, beaulieu2018photoexcitation, rozen2019controlling, comby2016relaxation} 
have been employed to discern enantiomers.
Here we demonstrate that TRXD within the pump-probe configuration offeres an alternative method to
probe the handedness of chiral molecules.

In previous work~\cite{hermann2020probing,tremblay2020time}, we demonstrated that the diffraction signal, obtained from TRXD, 
is related to the time derivative of the momentum-space density.
When a pump pulse induces a charge migration in matter,  
the time-dependent  charge dynamics is associated with the electronic flux densities by obeying 
the quantum continuity equation~\cite{sakurai1967advanced}.  
Time-resolved x-ray diffraction thereby encodes indirectly the information about electronic flux densities accompanying charge migration. 
The flux densities contains crucial mechanistic information about non-equilibrium electron dynamics in matter~\cite{okuyama2012dynamical, diestler2013computation,takatsuka2014chemical,hermann2014electronic, yamamoto2015electron,  bredtmann2014x,okuyama2009electron}. 
In recent years, several proposals have been put forward to image the flux densities and ring currents 
associated with charge migration using TRXD,
ultrafast x-ray absorption and high-harmonic generation~\cite{tremblay2020time, hermann2020probing, bredtmann2014x,popova2015imaging, carrascosa2020mapping,  nam2020monitoring, cohen19prl}. 
The present work shows that TRXD and electronic flux density analysis 
provide complementary information on the charge migration in oriented chiral molecules driven by linearly polarized light.
The intrinsic dynamics and its detection are drastically different, and the complementary information can thus be used to discriminate 
the enantiomers.

\section{Models and Methods}

We simulate ultrafast electron dynamics using the hybrid time-dependent density functional theory-configuration interaction (TDDFT-CI) methodology \cite{klinkusch2016,hermann2016tddft}.
This combination of methods provides a good balance between accuracy and computational efficiency.
A generic $N$-electron wave packet $\Psi(\mathbf{r}^N,t)$ is represented at any given time $t$ as 
a linear combination of the ground state Slater determinant and singly excited many-body excited states.
In spin-free representation, the wave packet takes the form
\begin{equation}\label{tdci}
\Psi(\mathbf{r}^N,t) = \sum\limits_{k=0}^{N_{\rm states}} C_k(t) \Phi_k(\mathbf{r}^N).
\end{equation}
In the case of field-free temporal evolution, $C_k(t)  = c_k~\textrm{exp}({-iE_{k}t})$. 
This wave packet evolves according to the many-electron time-dependent Schr\"{o}dinger equation 
\begin{equation}\label{eq01}
i \partial_t \Psi(\mathbf{r}^N,t) = \left[ \hat{H}_{0} - \hat{\mu} \cdot \mathcal{F}(t)\right]  \Psi(\mathbf{r}^N,t).
\end{equation}
Here, $\hat{H}_{0}$ is the field-free Hamiltonian within the Born-Oppenheimer approximation.
The field-molecule interaction is treated in the semi-classical dipole approximation,
with $\hat{\mu}$ the molecular dipole operator and $\mathcal{F}(t)$ an applied external electric field.

The lowest-lying excited eigenstates $\Phi_k(\mathbf{r}^N)$ of the many-body electronic Hamiltonian are calculated as linear combinations of Slater determinants.
Each excited $N$-electron state is expressed as a linear combination of singly excited configuration state functions
\begin{equation}\label{cis}
\Phi_k(\mathbf{r}^N) = \sum\limits_{ar} D_{a,k}^r \Phi_a^r(\mathbf{r}^N)
\end{equation}
The configuration state functions $\Phi_a^r(\mathbf{r}^N)$ are spin symmetrized combinations of excited Slater determinants.
They are defined with respect to the Kohn-Sham ground state $\Phi_0(\mathbf{r}^N)$,
by taking an electron from an occupied molecular orbital $a$ to a virtual molecular orbital $r$.

In the hybrid TDDFT-CI method,
all expansion coefficients $\{D_{0,k},D_{a,k}^r\}$ are obtained from linear-response TDDFT.
This step is performed using standard quantum chemistry programs.
The information from the quantum chemistry package is postprocessed using the open-source toolbox detCI@ORBKIT.
The contributions of the different Slater determinants extracted from the quantum chemistry programs are first pruned by removing all
contributions below some numerical threshold (here chosen as $10^{-5}$) and then renormalized.
In the basis of pseudo-CI eigenfunctions chosen for the propagation, the Hamiltonian is considered diagonal.
The matrix elements of the dipole moment operator in Eq.\,\eqref{eq01} are computed by numerical integration.
Finally, the time-evolution of the coefficients $C_k(t)$ in Eq.\,\eqref{tdci} is performed by direct numerical integration of Eq.\,\eqref{eq01}
using a preconditioned adaptive step size Runge-Kutta algorithm \cite{04:TC:passrk}.

In this work, all $N_\textrm{states}=31$ lowest-lying excited states below the ionization threshold are used to obtain convergence,
checked by repeating the laser-driven dynamics simulations with $N_\textrm{states}=\{16, 21, 26, 31\}$.
The states are computed using the CAM-B3LYP functional \cite{04:YTH:cam} and aug-cc-pVTZ basis sets \cite{dunning1989ccpvxz} on all atoms, 
as implemented in Gaussian16~\cite{frisch2016gaussian}. 
The motion of nuclei  are typically much slower in comparison to the electronic motion and 
therefore  nuclei are treated as frozen in this work.

Integrating the time-dependent many-electron wave packet in Eq.\,\eqref{eq01} leads to the one-electron quantum continuity equation
\begin{equation}\label{continuity}
 \partial_t\rho(\mathbf{r}, t) = -\vec{\nabla}\cdot \mathbf{j}(\mathbf{r},t).
\end{equation}
Here, $\rho(\mathbf{r}, t)$ is the one-electron density and $\mathbf{j}(\mathbf{r},t)$ the electronic flux density~\cite{barth2009concerted}.
From the many-body wavefunction, the expectation value of the one-electron density operator
\begin{eqnarray}\label{rho_op}
\hat{\rho} \left( \mathbf{r}\right) = \sum_{k}^{N} \delta\left( \mathbf{r} - \mathbf{r}_{k} \right)
\end{eqnarray}
yields the one-electron density.
The Dirac delta distribution  $\delta\left( \mathbf{r} - \mathbf{r}_{k} \right)$ evaluates the probability of finding electron $k$ with position
$\mathbf{r}_{k}$ at a point of observation,  $\mathbf{r}$.
The expectation value of the operator
\begin{eqnarray}\label{j_op}
\hat{j} \left( \mathbf{r}\right) = \frac{1}{2}\sum_{k}^{N} \left(\delta_k(\mathbf{r})\hat{p}_{k} + \hat{p}^{\dagger}_{k}\delta_k(\mathbf{r})\right)
\end{eqnarray}
yields the flux density, where the momentum operator $\hat{p}_{k}=-i\vec{\nabla}_{k}$ of electron $k$, 
is applied either to the right or to the left.
The molecular orbitals to compute the integrals are expressed as linear combinations of atom-centered Gaussian functions, and
all integrals are computed analytically 
The charge density and flux density thus obtained satisfy very accurately the continuity equation as given in Eq.\,\eqref{continuity}.
Details on the computation of the one-electron integrals from the many-body wavefunction using the detCI@ORBKIT
toolbox~\cite{hermann2016orbkit,pohl2017open, hermann2017open} are given elsewhere.

When the probe pulse duration is much shorter than the timescale of the charge migration, the
TRXD signal can be simulated using the expression for  the 
differential scattering probability (DSP)~\cite{dixit2014theory}
\begin{equation}\label{eq1}
\frac{dP}{d\Omega} =
\frac{dP_{e}}{d\Omega} \sum_{j} \left| 
\int \mathrm{d}\mathbf{r} ~\langle \Phi_{j}  | \hat{n}(\mathbf{r}) | \Psi({\textrm{T}}) \rangle~ e^{-i \mathbf{Q} \cdot \mathbf{r}}\right|^{2},
\end{equation}
where $\frac{dP_{e}}{d\Omega}$ is the Thomson scattering cross section of a free electron.
The signal probes the coherences between eigenstates of the unperturbed Hamiltonian, $| \Phi_{k}  \rangle$,
contained in an electronic wave packet $| \Psi({\textrm{T}}) \rangle$ with ${\textrm{T}}$ the pump-probe delay time. In addition,
$\mathbf{Q}$ is the photon momentum transfer and $\hat{n}(\mathbf{r})$ is the electron density operator.
The summation over $j$ includes all zeroth-order states included in the dynamics, which amounts
to detecting all the scattered photons within a given solid angle, irrespective of the mean photon energy.

After the pump pulse 
and in the absence of an external field, the time-dependent coefficients in Eq.\,\eqref{tdci} simplify
to the analytical form containing an amplitude and energy-dependent phase.
At ${\textrm{T}} = 0$, Eq.\,\eqref{eq1} reduces to the time-independent diffraction signal as 
\begin{equation}\label{eq3}
\frac{dP}{d\Omega} = 
\frac{dP_{e}}{d\Omega} \sum_{j}  \left|  \sum_{k} c_{k} 
\int d\mathbf{r} ~\langle \Phi_{j}  | \hat{n}(\mathbf{r}) | \Phi_{k} \rangle~ e^{-i \mathbf{Q} \cdot \mathbf{r}}\right|^{2}. 
\end{equation}
As evident from this equation, 
the time-independent diffraction signal is emerging from the 
joint contribution of the elastic and time-independent inelastic parts 
of the total diffraction signal. This term 
serves as a constant background in TRXD. 
To image the charge migration, we will present  
time-dependent difference diffraction signals 
where the total diffraction signal at zero delay time is subtracted 
from the subsequent delay times. 

References\,\cite{kowalewski2017monitoring, simmermacher2019theory} discuss the different 
roles of time-independent and time-dependent diffraction signals  from the total diffraction signal. Moreover, detailed discussions about probing electronic coherences in the 
 wavepacket have been documented in Refs.~\cite{dixit2012, simmermacher2019electronic}. 
 Furthermore, the time resolution in TRXD depends on several factors such as temporal duration on x-ray pulse, jitter between pump and probe pulses, and pump-probe time delay.

\begin{figure*}[htb]
\includegraphics[width = 0.9 \linewidth]{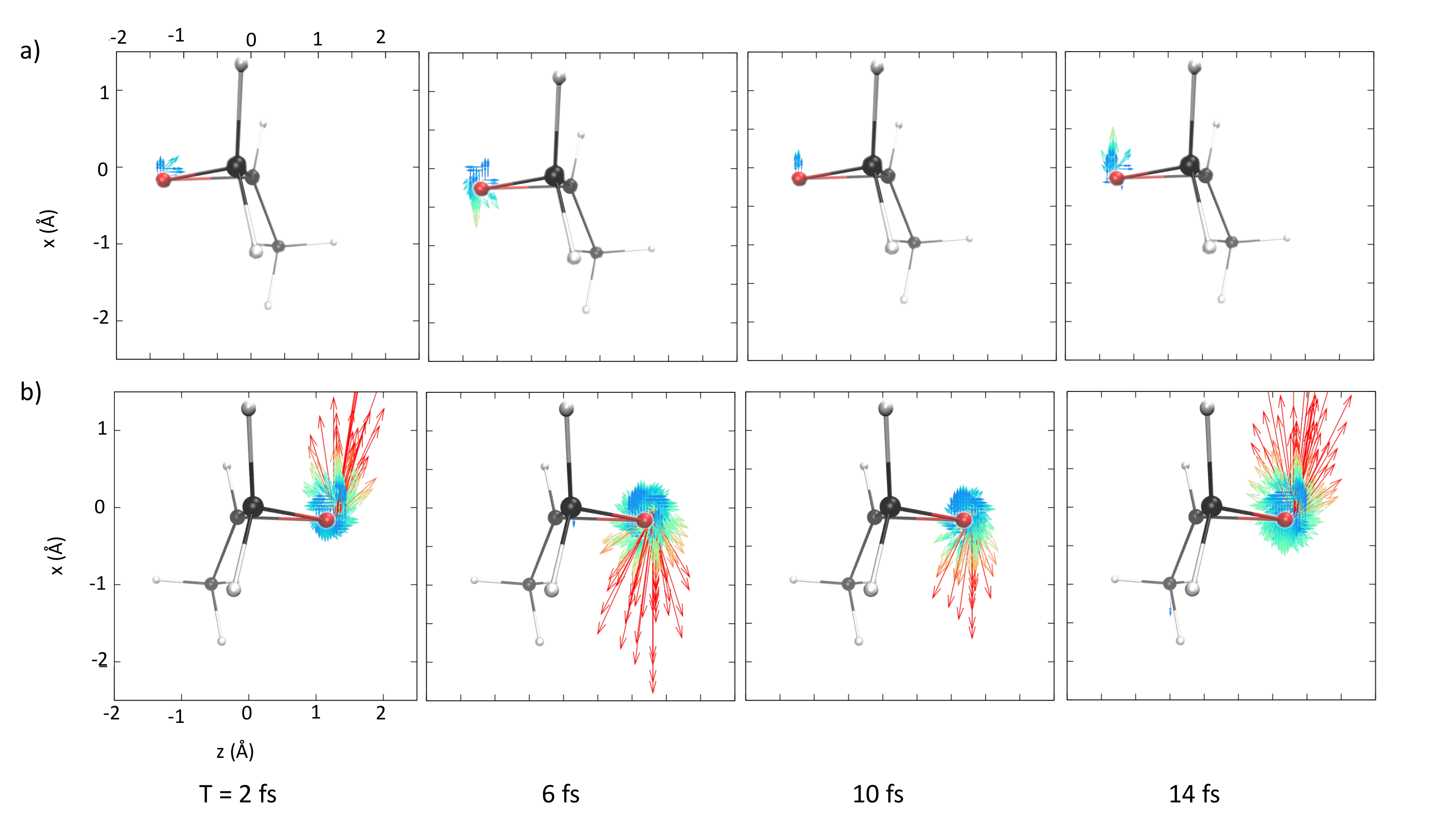}
\caption{Time-dependent electronic flux densities for (a) R- and, (b) S-enantiomers of epoxypropane in the $xz$-plane at $2$ fs, $6$ fs, $10$ fs, and $14$ fs
after the pump pulse.
Violet, maroon and gray colors are representing carbon, oxygen and hydrogen atoms, respectively. 
Here, the $z$-axis is defined as the chiral axis in the molecular frame of reference.} \label{fig01}
\end{figure*}

All many-electron dynamical simulations are performed using in-house codes \cite{klinkusch2011,04:TC:passrk,08:TKS,08:TS:gloct,10:TKKS}.
The integrals required to compute the TRXD signals and electronic flux densities are calculated using ORBKIT~\cite{OKgit}.
Mayavi \cite{mayavi} and Matplotlib \cite{hunter2007matplotlib} are used to visualize the electronic flux densities 
and the time-resolved x-ray signals, respectively.

\section{Results and Discussion}
Our starting point is a pair of oriented R- and S-enantiomers of epoxypropane.
In recent years, it has become possible to orient chiral molecules in space. 
This has been demonstrated both theoretically and  experimentally in Refs.~\cite{tutunikov2018selective, tutunnikov2020observation, saribal2020detecting}.
Orientation can be achieved using either skewed-polarized pulses, a centrifuge, or a combination of a static field and linearly polarized pulse.
The resulting orientation has the R- and S-enantiomers pointing in opposite directions along the chiral axis. 
As the molecules are fixed in their own frame of reference, everything in this work is described in the molecular frame,
where $z$ defines the chiral axis.

As in previous work, oriented epoxypropane is exposed to a linearly polarized pump pulse of 10\,fs pulse duration with 9.1$\times$10$^{14}$ W/cm$^{2}$ peak intensity to initiate the charge migration dynamics. The pulse is polarized at 45$^\circ$ in the $yz$-plane of the molecular frame of reference and has a 160\,nm wavelength with sine-squared envelope. 
As a result of the interaction between the pump pulse and epoxypropane, ultrafast charge migration is set into motion.

For the R-enantiomer, $32\%$, $7\%$, and $2\%$ electronic populations are transferred by the end of the pump pulse from ground state 
to the 3$^{\text{rd}}$, 4$^{\text{th}}$ and 5$^{\text{th}}$ excited electronic states, respectively.
The rest of the population remains in the ground state, which is responsible for a dynamics on the attosecond timescale.
On the other hand, the 3$^{\text{rd}}$, 4$^{\text{th}}$ and 5$^{\text{th}}$ excited states are populated at
$37\%$, $10\%$ and $13\%$, respectively, while nearly 30$\%$ remains in the ground state for the S-enantiomer~\cite{giri2020time}. 
The reason behind different electronic populations in R- and S-enantiomer can be attributed to different temporal evolution of the magnitude and the phase of the Cartesian components of the 
total dipole moment, as discussed in Ref.~\cite{giri2020time}.
Since the populations at the end of the pump pulse are different for both enantiomers,
the field-free charge migration looks significantly different in both cases. This can be revealed by 
investigation of the transient flux densities.

The electronic flux densities accompanying the charge migration for both enantiomers are presented in Fig.~\ref{fig01}. 
To avoid the fast oscillating contribution stemming from the 
coherent interaction with the ground state in the charge migration, 
only contributions from the excited states are considered in the reconstructed signal.
Snapshots are presented for $2$ fs, $6$ fs, $10$ fs, and $14$ fs, and the 2D representation in the $xz$-plane is obtained
by integrating over the $y$ coordinate.
The arrows of the vector field are color-coded according to their magnitude.
In both enantiomers, almost all the flux density is concentrated around the oxygen atom (in maroon).
It is important to mention that, maybe surprisingly, the chiral carbon atom is not involved in the dynamics,
as revealed by the transient flux densities~\cite{giri2020time}. 
The most striking difference lies in the magnitude of the currents.
The strength of the flux densities for the R-enantiomer is weak in comparison  to that for the S-enantiomer during the charge migration. 
At all times, they both exhibit a similar circular pattern around the oxygen atom.
The direction of the circular charge migration changes as the dynamics progresses.

Since the direction of the electron flow is more prominent for the S-enantiomer, as reflected in Fig.~\ref{fig01}(b),
it can therefore be more readily described.
Arrows corresponding to the flux densities are emerging from the oxygen atom in all directions,
corresponding to an approximately circular current density.
Yet, the longest arrows in red represent the most probable direction for the flow of electrons,
and these are changing with time.  
At $2$ fs, arrows are mostly emerging in the positive $x$-direction and it  is almost reversed in the next snapshot, i.e., at $6$ fs.
The flux densities for the S-enantiomer are along the same direction at $2$ and  $14$ fs, which indicate  
the partial recurrence. 
Similar observations can be made for the circular component of the current density in the R-enantiomer, where the 
direction of the circular current is opposite to that of the S-enantiomer.
In contrast, the current component with the largest magnitude points in the same direction for both enantiomers at all times.
A tentative explanation could be that the circular component stems from the phase transferred to the molecule by the laser field. 
This leads to opposite relative phases among the three excited states.
On the other hand, the dominant component of the flux density could be associated with the response of the electron cloud
in the electric field during the preparation phase. This electronic response behavior would then be inherited by the subsequent
charge migration dynamics, explaining that both enantiomers react with the same phase.

The different timescale of the charge migration can be understood by analyzing 
the characteristic timescales  involved in the system, i.e., $\tau = \textit{h} / \Delta E$, where $\Delta E$ 
is the energy difference between two eigenstates. The characteristic timescales corresponding to 
$E_3$ = 7.56 eV, $E_4$ = 7.73 eV, and $E_5$ = 7.84 eV are $\tau_{43} = 24.32$ fs, $\tau_{53} = 14.77$ fs and $\tau_{54} = 37.59$ fs.
Different contributions of the excited states and their respective interference including these timescales are the reason for non-identical charge migration
in the two excited enantiomers.
\begin{figure}[ht!]
\includegraphics[width=1.0\linewidth]{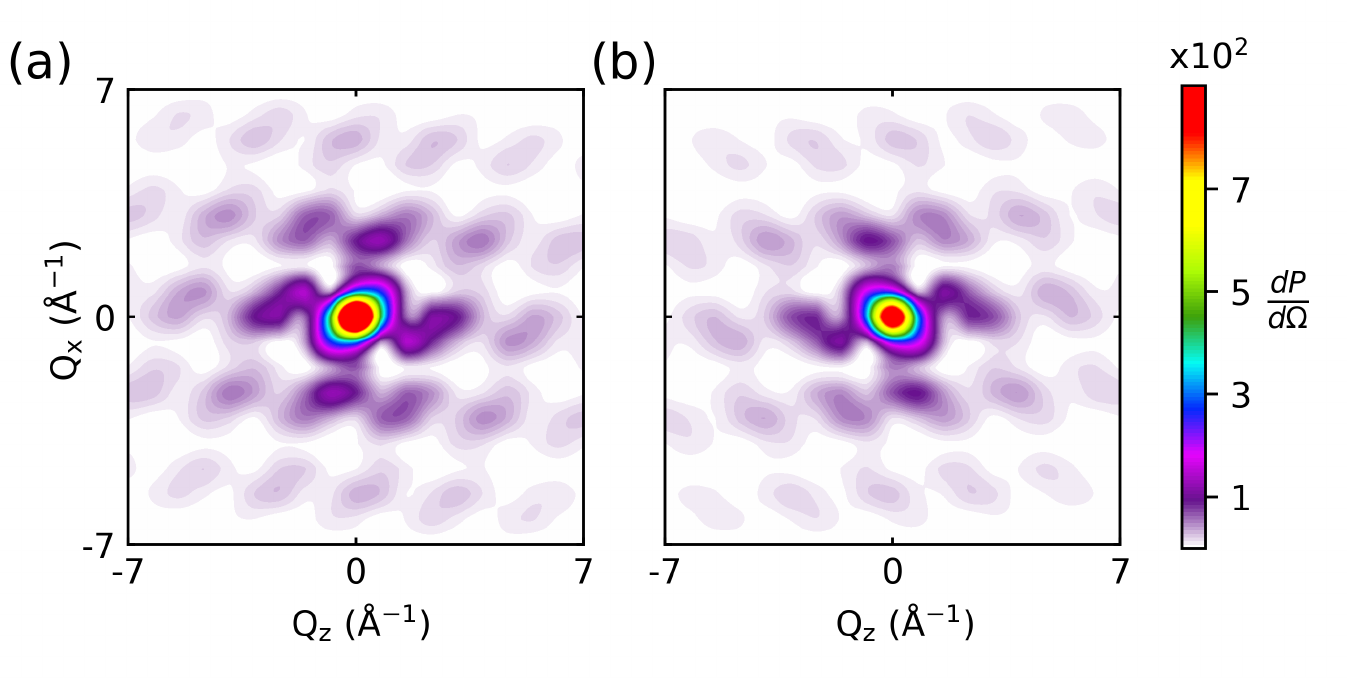}
\caption{Time-independent diffraction patterns for  (a) R- and (b) S- enantiomers of epoxypropane. The  diffraction patterns are presented in  $Q_{z}-Q_{x}$ plane in units of  
${dP_{e}}/{d \Omega}$. An ultrashort x-ray pulse of 8 keV mean energy is aimed to obtain 1.55~\AA~spatial resolution in these simulations. 
All diffracted photons up to 60$^\circ$ are collected in the detector.} \label{fig02}
\end{figure}

Time-resolved x-ray diffraction is a technique that can be used  to image the dynamics  of charge migration with atomic-scale spatiotemporal resolutions. In recent years, different formalisms of TRXD have been developed to image an increasing number of  
processes~\cite{henriksen, dixit2013jcp, bennett2018monitoring,  simmermacher2019electronic, dixit2013prl, dixit2014theory, santra2014comment, bredtmann2014x, dixit2017time, kowalewski2017monitoring}.
To simulate time-resolved diffraction signal,  the DSP  is computed according to Eq.\eqref{eq3}.
The six lowest-lying excited eigenstates (i.e., $j = [0, 6]$) were sufficient to obtain the converged diffraction signals under the assumption that the sum-over-states expression was truncated due to the detector response.

\begin{figure*}[]
\includegraphics[width=0.9\linewidth]{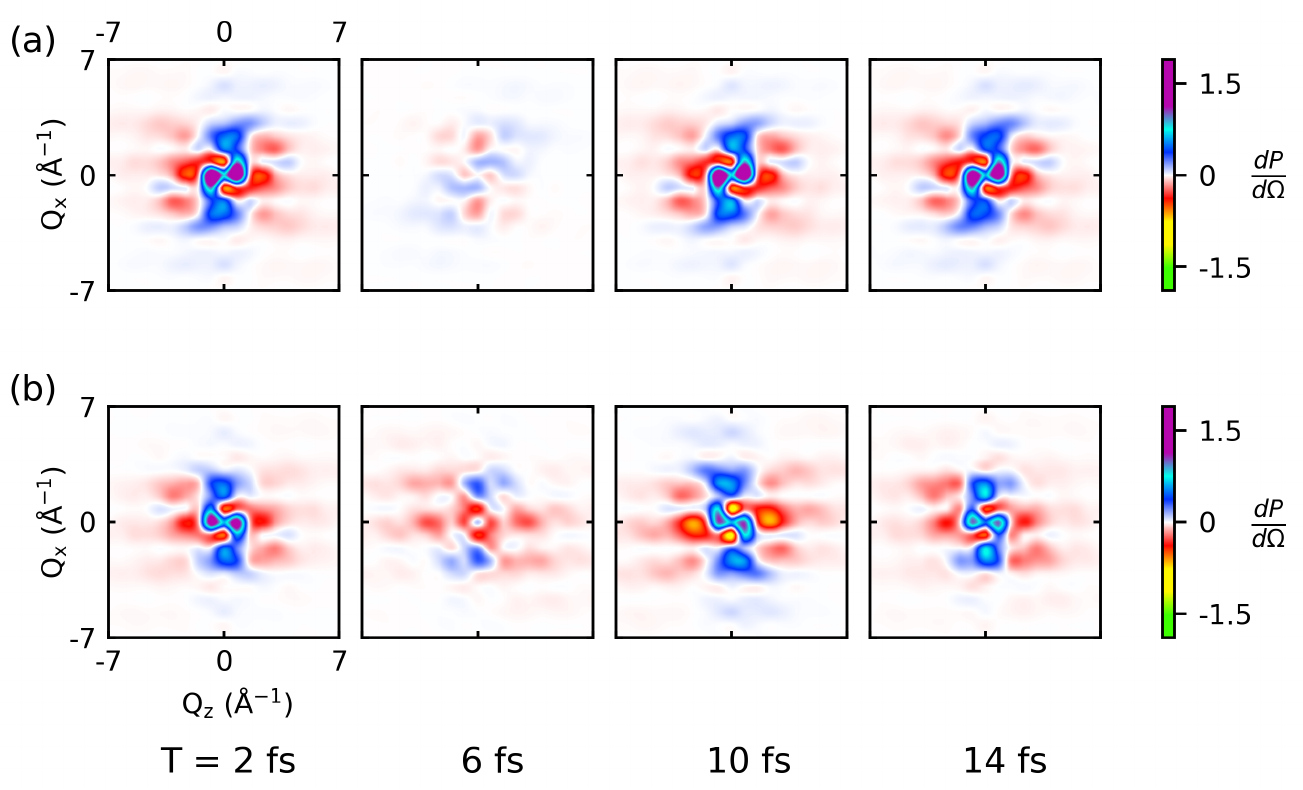}
\caption{Time-resolved diffraction patterns for (a) R- and, 
(b) S-enantiomers of epoxypropane 
in $Q_{z}-Q_{x}$ plane at  2 fs, 6 fs, 10 fs, and 14 fs. 
Intensity of the diffraction patterns is presented in units of  ${dP_{e}}/{d \Omega}$. 
Here, time-dependent diffraction signal at T = 0 fs is subtracted 
to the subsequent delay time.  An ultrashort x-ray pulse of 8 keV mean energy is aimed to obtain 1.55~\AA~spatial resolution in these simulations. 
All diffracted photons up to 60$^\circ$ are collected in the detector.} \label{fig03}
\end{figure*}

The time-independent contributions to the total diffraction signal for both enantiomers after the excitation are presented in Fig.~\ref{fig02}. 
 These signals are simulated using Eq.~(\ref{eq3}). 
From the figure, one can see that the time-independent diffraction signals 
are different for both enantiomers. First, the intensity of  the 
time-independent  signal  is stronger for the R-enantiomer in comparison to the S-enantiomer.   This is especially true in the low-momentum region.
Also, the signals are not mirror images of each other, which could allow one to discriminate between the enantiomers by simple inspection of the two diffraction signals.
Since these diffraction signals 
are static in nature, the dynamical information about the charge migration is lost.

Time-resolved diffraction signals at selected delay times  for both enantiomers are presented in Fig~\ref{fig03}.
The total diffraction signal at T = 0 fs is subtracted from the signal at subsequent delay times.
It is observed that the time-independent signal is three orders of magnitude stronger than this time-dependent signal. 
Similar observations have been reported earlier in the case of TRXD for different systems~\cite{carrascosa2020mapping, keefer2021}.  
In spite of having low magnitude, the information about the ultrafast dynamics of charge migration 
is imprinted  in the different  diffraction signals as they are significantly different for 
 both enantiomers at different delay times. 
The central parts as well as the higher-momentum signals 
are changing in a very distinct way for both enantiomers. Note that $\langle \Phi_{i} \vert \Phi_{j} \rangle$ is not exactly zero in our 
case due to numerical limitations, which impose using a small but finite value with a threshold value of  $10^{-5}$ on the linear-response TDDFT coefficients. 
This, together with the numerical coefficient renormalization, explains the small but non zero value of the diffraction signal at the center in Fig.~\ref{fig03}.

\begin{figure*}[ht!]
\includegraphics[width=0.9\linewidth]{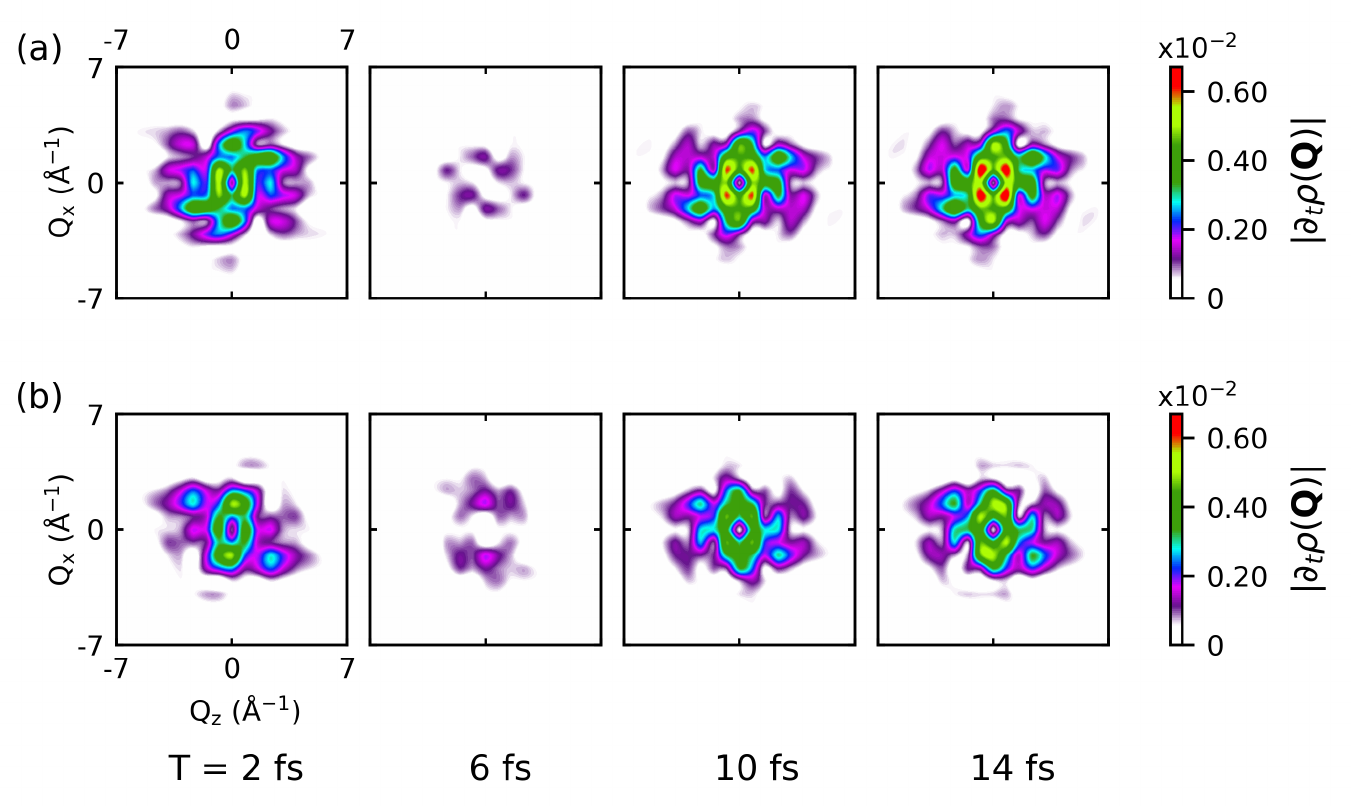}
\caption{Time derivative of electron density in momentum space, 
$\vert \partial_t\rho({\mathbf{Q}})\vert$, for  (a) R- and, b) S-enantiomers  
of epoxypropane in $Q_{z}-Q_{x}$ plane at 2 fs, 6 fs, 10 fs, and 14 fs.} \label{fig04}
\end{figure*}

By comparing the time-dependent diffraction pattern of the enantiomers at 2 fs, it can be seen that the overall signals are not very different for both enantiomers.
There are two lobes of positive signal around the center and they are of almost the same intensity with different orientations for the respective enantiomers.
The negative part of the signal is scattered at higher momenta. There is a mismatch in the intensity at the central part of the lobes for the two enantiomers.
 
At the next time step, i.e., at 6 fs, the diffraction signal is reduced significantly for both enantiomers.
The signal is just reversed around the center, dominated by a negative value  and spreading to higher momenta for the S-enantiomer. 
For the R-enantiomer, the signal at 10 and 14 fs are similar to the one at 2 fs and reflect slower dynamics. In contrast, for the S-enantiomers, the signal at 10 fs is comparable to the signal at 2 fs but not identical due to the presence of four dips with a minimum value around the center. There is overall depletion of the signal at 14 fs compared to the previous time step.
 The oscillation period of the time-dependent signal correlates
well with the timescale of the charge migration found in the flux density for the enantiomers as shown in Fig.~\ref{fig01}.

In order to understand how the time-dependent diffraction signals are related to 
the flux density, the
time derivative of the electron density in momentum space,  
$\vert \partial_t\rho({\mathbf{Q}})\vert$, at selected delay times is presented  in Fig.~\ref{fig04}. At T = 0 fs, $\vert \partial_t\rho({\mathbf{Q}})\vert$ is zero  (see Eq. (S7) in Ref.~\cite{hermann2020probing}). Therefore, subtracting $\vert \partial_t\rho({\mathbf{Q}})\vert$ 
at T = 0 fs is not required. The
$\vert \partial_t\rho({\mathbf{Q}})\vert$ is related to the flux densities, and it 
can be obtained by a Fourier transform of the electronic continuity equation, Eq.\,\eqref{continuity}.

For both enantiomers, the overall signal of $\vert \partial_t\rho({\mathbf{Q}})\vert$ decreases drastically from 2 fs to 6 fs. 
There is a partial revival of the signal at 10 fs for the R-enantiomer as observed in Figs.~\ref{fig01} and ~\ref{fig03}. 
The intensity of $\vert \partial_t\rho({\mathbf{Q}})\vert$  at 14 fs increases compared to that in 10 fs for the R-enantiomer. In contrast, $\vert \partial_t\rho({\mathbf{Q}})\vert$  for the S-enantiomer at 14 fs is approximately similar to the one at 2 fs.

As evident from Figs~\ref{fig03} and \ref{fig04}, it is not straightforward to establish visual one-to-one  
correspondence between the DSP and $\vert \partial_t\rho({\mathbf{Q}}) \vert$, which is in contrast to the
earlier work discussed in Ref.~\cite{hermann2020probing}. 
There are two simple reasons behind this. The first one is that more than two electronic states 
are contributing to the electronic wavepacket in epoxypropane. 
As a result, time-dependent terms yield more than one sine term in the expression of 
$\vert \partial_t\rho({\mathbf{Q}}) \vert$ (see Eq. (S7) in Ref.~\cite{hermann2020probing}). 
The other reason can be attributed to a significant contribution stemming  from the sine square term in comparison to the sine  term in the expression of $ \Delta {dP}/{d \Omega}$ (see Eq. (S12) in Ref.~\cite{hermann2020probing}). These findings are in contrast to an earlier case, 
where the wavepacket was composed of only two electronic states,
and the sine square term was negligibly small in comparison to the sin term~\cite{hermann2020probing}. 
Therefore, the comparison between the DSP and $\vert \partial_t\rho({\mathbf{Q}})\vert$ can only be qualitative.

\begin{figure*}[bht!]
\includegraphics[width=0.7\linewidth]{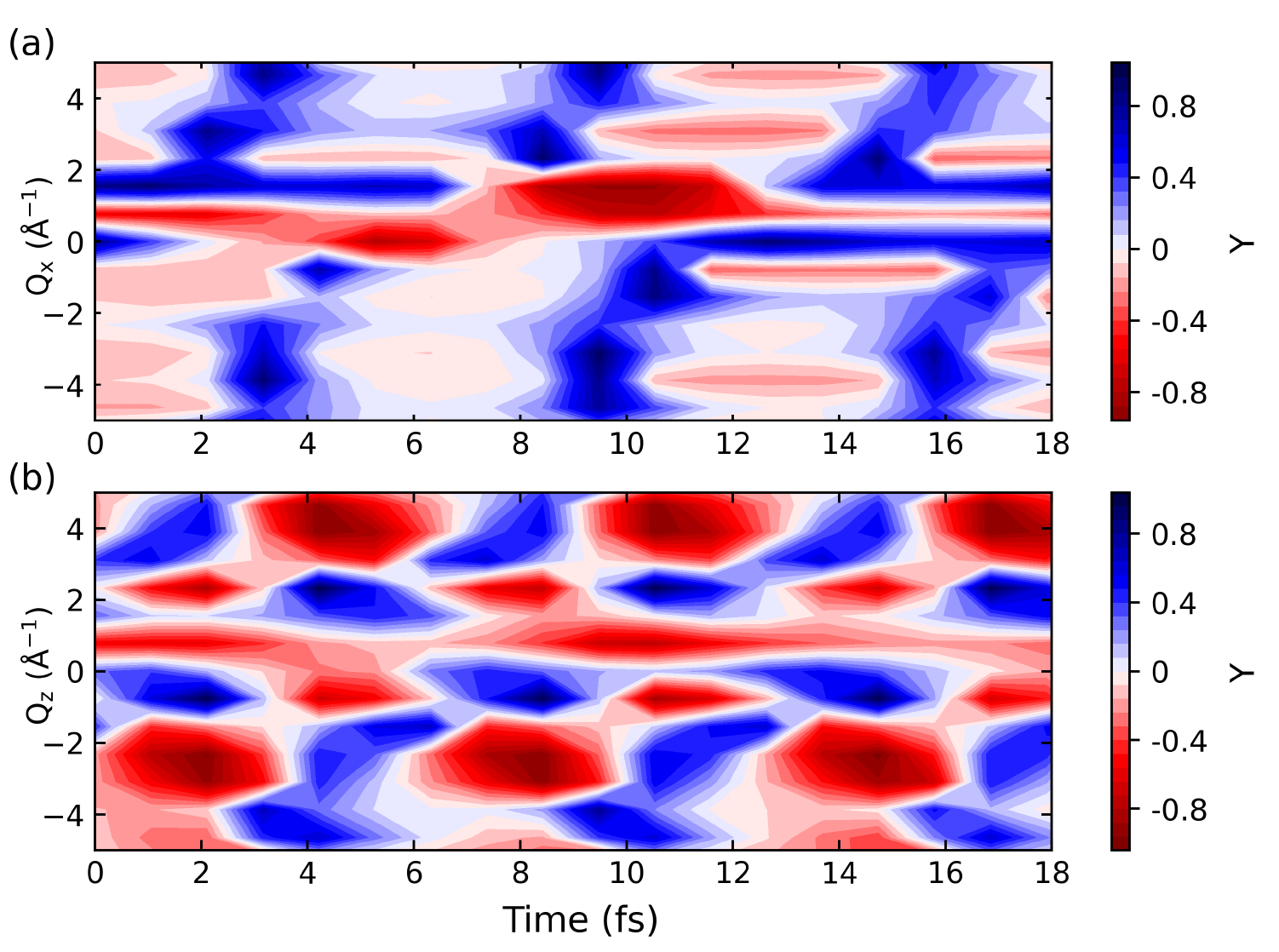}
\caption{Asymmetry parameter $ Y $ along the (a) $Q_{x}$ and, (b) $Q_{z}$ axes as a function of  the delay time. The 1D TRXD signal is obtained from Eq.\,\eqref{eq1} and integrated along the other axes.} \label{fig05}
\end{figure*}

As the time-resolved diffraction signals for the two enantiomers are different, 
we can define the asymmetry parameter  as 
\begin{equation}\label{eq4} 
Y =  \frac{\sigma_R - \sigma_S}{\sigma_R + \sigma_S}, 
\end{equation} 
where $\sigma_R$ and $\sigma_S$ are the DSPs  of the R- and S-enantiomers, respectively. 
Here, the DSP is represented by $\sigma$ as an abbreviation. 
Figures~\ref{fig05}(a)  and ~\ref{fig05}(b) present, respectively,  
the asymmetry parameter  along $Q_{x}$ and $Q_{z}$ axes, respectively, as a function of time.
The domination of the red part over the blue is an indication that the signal for the S-enantiomer is stronger than that for the R-enantiomer along the $Q_{x}$ axis. 
If we look only around the $Q_{x} = 0$ line, the signal for the R-enantiomer is dominating over the S-enantiomer for all the time steps. The case is almost reversed for $Q_{x} \neq 0$.  

The signals are more interesting along the chiral axis, i.e., along the $Q_{z}$ axis. 
For a specific point in momentum space, the signal is changing in a periodic fashion. 
The same is true if we fix the time-delay and walk along the $Q_{z}$ axis. 
These signals encode important information about the dynamical evolution of the system, which
is unique for a given choice of laser tagging parameters.
The asymmetry parameter can be used to determine the enantiomeric ratio in a sample of
unknown concentrations $C_R$ and $C_S$ in the R and S enantiomers, respectively.
In this case, the measured signal of the unknown mixture is given as the weighted sum of the two signals- 
$\sigma_\textrm{sample} = C_R\sigma_R + C_S\sigma_S$.
By subtracting from the TRXD signal of a pure enantiomer, say, $\sigma_R$, this difference
\begin{equation}\label{eq5} \begin{aligned}
\Delta\sigma &=  \sigma_R - (C_R\sigma_R + C_S\sigma_S)\\
&= (1-C_R)\sigma_R - C_S\sigma_S= C_S(\sigma_R - \sigma_S)\\
\end{aligned}\end{equation} 
can be simply normalized and related to the asymmetry parameter
\begin{equation}\label{eq6} 
\frac{\Delta\sigma}{\sigma_R + \sigma_S}
=  C_S\left(\frac{\sigma_R - \sigma_S}{\sigma_R + \sigma_S}\right) = C_SY, 
\end{equation} 
Hence, from the knowledge of the TRXD signals of both enantiomers, 
the concentration of an enantiomer can be directly obtained by comparing the intensity of the difference
signal with that of the asymmetry parameter. As the dynamics becomes more intricate at longer times,
the time-dependent monitoring of the different signals would potentially lead to a more 
precise determination of the concentrations.
  
\section{Conclusion}  
The present work aimed at imaging the charge migration in epoxypropane using TRXD. 
The charge migration in oriented epoxypropane was triggered by a linearly polarized pulse  at 45$^\circ$ in the 
$yz$-plane in the molecular frame of reference.  
For each enantiomer,  the induced charge migration was different, and  is imaged by TRXD. 
The dominating time-independent  and different TRXD signals 
were analyzed separately.
It was found that time-dependent diffraction signals are significantly different for both enantiomers.
We believe  that the present proof-of-principle results should motivate further studies of laser-induced charge migration
in chiral molecules imaged by TRXD. 
In particular, this could pave the way for a new, sensitive technique to determine the enantiomeric concentrations
in unknown samples.

\section*{Acknowledgements}
G. D. acknowledges support from Science and Engineering Research Board of India 
(Project No. ECR/2017/001460) and the Ramanujan Fellowship (Grant No. SB/S2/ RJN-152/2015). S. G. acknowledges CSIR for a Senior Reasearch Fellowship. We are grateful to the anonymous referee for constructive suggestions.  


\end{document}